\documentclass{article}
\usepackage{ijcai19}

\usepackage{times}
\usepackage{soul}
\usepackage{url}
\usepackage[hidelinks]{hyperref}
\usepackage[utf8]{inputenc}
\usepackage[small]{caption}
\usepackage{graphicx}
\usepackage{amsmath}
\usepackage{booktabs}
\usepackage{amssymb,amsfonts}
\usepackage{bm}

\usepackage{theorem}

\newtheorem{theorem}{Theorem}
\newtheorem{definition}{Definition}
\newtheorem{lemma}{Lemma}

\usepackage{algorithm}
\usepackage[noend]{algorithmic}
\usepackage{eqparbox}
\usepackage{color}

\newcommand{\W}{\mathsf{W}}
\newcommand{\F}{\mathsf{F}}

\newcommand{\cI}{\mathcal{I}}
\newcommand{\cP}{\mathcal{P}}
\newcommand{\bR}{\mathbb{R}}

\newcommand{\items}{\mathcal{M}}
\newcommand{\valu}{V}
\newcommand{\MMS}{\mathsf{MMS}}
\newcommand{\WMMS}{\mathsf{WMMS}}
\newcommand{\OWMMS}{\mathsf{OWMMS}}

\newcommand{\share}{s}

\DeclareMathOperator*{\argmax}{arg\,max}

\newcommand{\qed}{\unskip\hspace*{1em}\hspace{\fill}$\Box$}
\newenvironment{proof}[1][Proof]{\begin{trivlist}
  \item[\hskip \labelsep {\it #1:}]}{%
    \qed\end{trivlist}}

\title{Weighted Maxmin Fair Share Allocation of Indivisible Chores}

\author{
Haris Aziz$^1$\and
Hau Chan$^2$\And
Bo Li$^3$\\
\affiliations
$^1$UNSW Sydney and Data61 CSIRO, Australia\\
$^2$Department of Computer Science and Engineering, University of Nebraska-Lincoln, USA\\
$^3$Department of Computer Science, Stony Brook University, USA\\
\emails
haziz@cse.unsw.edu.au,
hchan3@unl.edu,
boli2@cs.stonybrook.edu
}

\begin{document}

\maketitle

\begin{abstract}
We initiate the study of indivisible chore allocation for agents with asymmetric shares.
The fairness concept we focus on is the weighted natural generalization of maxmin share: WMMS fairness and OWMMS fairness. 
We first highlight the fact that commonly-used algorithms that work well for the allocation of goods to asymmetric agents, and even for chores to symmetric agents do not provide good approximations for allocation of chores to asymmetric agents under WMMS. 
As a consequence, we present a novel polynomial-time constant-approximation algorithm, via linear program, for OWMMS.
For two special cases: the binary valuation case and the 2-agent case, we provide exact or better constant-approximation algorithms.
\end{abstract}

\section{Introduction}

We consider fair allocation of indivisible chores when agents have asymmetric shares. In contrast to the case of goods for which agents have positive value, chores are disliked by agents and they have negative values for them. The fairness concept we focus on is the maxmin share (MMS) fairness which was designed for allocation of indivisible items. MMS is based on the thought experiment that if the items are partitioned into bundles and an agent would always get the least preferred bundle of items, what is the best way she can partition the items. The value of such a bundle is the maxmin share of the agent. An allocation is deemed MMS fair if each agent gets her required share. 

Maxmin share fairness was proposed by \cite{Budi11a} as a fairness concept for allocation of indivisible items. It is a relaxation of \emph{proportionality} fairness that requires each of the $n$ agents should get a value that is at least $1/n$ of the total value she has for the set of all items. When items are divisible, maxmin share fairness coincides with proportionality. 
Maxmin share fairness is a weaker concept when items are indivisible. 
It was conjectured that a maxmin fair allocation always exists but \cite{PrWa14a} identified a counter-example. Since the work of \cite{PrWa14a}, there are several papers on algorithms that find an approximate MMS allocation~\cite{AMNS15a,BM17a,SGHSY18,ARSW17a}. All these works make a typical assumption that agents are symmetric and should be treated in a similar manner.

\cite{FHG+17a} were the first to consider MMS fairness for the case where indivisible goods are allocated and the agents are \emph{not} symmetric because they may have different entitlement share of the goods. 
Ideally, an agent would expect to get a share of the total value that is proportional to her entitlement. However, when items are indivisible, MMS fairness needs to be suitably generalized to the cater for asymmetric entitlement shares. \citeauthor{FHG+17a} generalized MMS fairness to that of the more general MMS concept as \emph{weighted MMS (WMMS)} that caters for entitlements.
They devised a simple ordinal (that only used the qualitative ranking information of items) algorithm that ensures an 
$n$-approximation guarantee for WMMS where each agent's allocation is at least $1/n$ of her value in a WMMS allocation. 
Beyond the results for goods~\cite{FHG+17a,FHG+19a}, not much is known about chore allocation when the agents are asymmetric despite the recent active research in fair allocation of goods and chores. Furthermore, it is not clear whether the results for goods from one setting could carry over the other~\cite{Aziz16a}.

In this paper, we focus on the fair allocation of \emph{chores} rather than goods for asymmetric agents. 
In the case of chores, agents do not have entitlements but relative shares. If an agent has a higher share, she is expected to take a higher load of the chores. Treating agents asymmetrically may be a requirement for several reasons. For example, countries with a larger population and CO2 emission may be liable to undertake more responsibility to clean up the environment. In this paper, the central research question we examine is the following one.
\emph{When indivisible chores are to be allocated among agents with asymmetric shares, for what approximation factor do approximately WMMS fair allocations exist and how efficiently can they be computed?}

\subsection*{Contributions}
We consider a model of allocation of chores in which agents have relative shares as compared to entitlements. 
Different to the case of symmetric agents, we first prove that even with only two agents,
no algorithm can simultaneously guarantee each agent's value to be higher than $\frac{4}{3}$ of her weighted maxmin share.
Moreover, we show that many greedy algorithms widely used in the literature, including \cite{FHG+17a} and \cite{ARSW17a},
may have arbitrarily bad performance.

Then we design a polynomial-time algorithm which
provides a 4-approximation to the minimal relaxation of WMMS value (OWMMS) under which a WMMS allocation exists.
To present this algorithm, we first study a special case when all agents have an identical valuation.
The algorithm combines (1) the use of a greedy algorithm for the case of identical valuations and (2) linear programming and rounding techniques.

Finally, we study two restricted cases: a two-agent setting and a binary valuation setting.
For the two-agent case, we present a variant of divide-and-choose protocol which
ensures each agent's value is at least $\frac{3}{2}$ of her weighted maxmin share;
For binary valuations, we show that a WMMS allocation exists and can be efficiently computed. For asymmetric agents and indivisible items, this is the first algorithmic result for binary valuations. 

\section{Related Work}

The fair allocation problem has been extensively studied in the cake cutting literature
\cite{dubins1961cut,stromquist1980cut,alon1987splitting,brams1995envy,brams1996fair,robertson1998cake,aziz2016discreten}.
In this line of work, researchers study how to fairly allocate a divisible item (e.g., cake)
among a number of agents. Solution concepts such as envy-freeness and proportionality
are prominent criteria for fairness. In the context of divisible goods, researchers have extended results for the case of equal entitlements to those of unequal entitlements~(see e.g., \cite{CF18a}). 

As for MMS fairness, it is already known that even for additive valuations, there exists
an instance such that no allocation can simultaneously guarantee each agent receives at least her MMS \cite{kurokawa2018fair}.
But approximate MMS can be efficiently computed; see \cite{BM17a,SGHSY18,kurokawa2018fair}. Computing WMMS shares is an NP-hard problem for both goods and for chores even for the case of 2 agents and for equal shares. The statement can be derived via a reduction from the integer partition problem~\cite{GaJo79a}.

Most of the work on fair allocation of items is for the case of goods although recently,
fair allocation of chores~\cite{ARSW17a} or combinations of goods and chores~\cite{ACI19a} has received attention as well.
It is shown by \cite{ARSW17a} that MMS allocations for chores do not always exist but can be 2-approximated
by a simple round-robin algorithm. \cite{ARSW17a} also presented a PTAS for relaxation of MMS called optimal MMS.
\cite{BM17a} presented an improved approximation algorithm for MMS allocation of chores. 
Fair allocation of indivisible goods and asymmetric agents has also been studied~\cite{FHG+17a,FHG+19a}.
We take a similar approach and study the \emph{chore} allocation problem when the agents are not symmetric. 
\cite{BNT17a} considered the allocation of indivisible goods where agents have different entitlements. One of the concepts that they propose is called $\ell$-out-of-d MMS that can also apply to agents having ordinal preferences over bundles of chores. However, the paper focusses on results for goods. 

\section{Preliminaries}
We begin by presenting our setting formally
and discussing fairness concepts
as well as some basic notations in the paper. 

\subsection{Setting}
Let $N=\{1,2,\cdots,n\}$ be a set of $n$ agents, and $M=\{1,2,\cdots,m\}$ be a set of $m$ indivisible items.
In this work, we always use $i\in N$ and $j\in M$ to indicate an agent and an item, separately.
Each agent has a valuation function $\valu_i: 2^{M}\to \mathbb{R}$.
Denote by $V_{ij}=V_i(\{j\})$.
We assume that items are chores to every agent, i.e., $V_{ij}\leq 0$ for all $j\in M$ and 
the valuations are additive, i.e., for any $S\subseteq M$, $V_i(S)=\sum_{j\in S}V_{ij}$. 
Without loss of generality and just for ease of presentation, throughout this paper except Section \ref{sec:binary},
it is assumed that all of the valuations are normalized, i.e. $V_{i}(\emptyset)=0$ and $V_{i}(M)=-1$. 

In this work, we consider the case when agents are asymmetric.
Particularly, every agent has a share for the chores, namely $\share_i \in (0, 1]$.
The shares add up to $1$, i.e., $\sum_{i\in N} \share_i = 1$.

Letting $\bm{V}=(V_{1},\cdots,V_{n})$ and $\bm{s}=(s_{1},\cdots,s_{n})$,
we use $\cI=(N,M,\bm{s},\bm{V})$ to denote a chore allocation instance and
$\cI=(N,M,\bm{s},V)$ when all agents have the identical valuation $V$.  
Note that when all agents have identical valuation $V$, $V(\{j\})$ is simplified as $V^{j}$ for any $j\in M$.
Let $\Pi(M)$ be the set of all $n$-partitions of the items.
A generic allocation will be denoted by  $X= \langle X_1, \ldots, X_n\rangle$ where $X_i$ is the bundle of agent $i$.

\subsection{WMMS Fairness}

Before presenting the WMMS fairness concept that takes into account the shares of the agents, we first present the standard MMS fairness concept that assumes the shares of the agents are equal.
For symmetric agents, the classical {\em maxmin share} (MMS) of an agent $i$ with valuation $V_{i}$ is defined as
$$
\MMS_i = \max_{\langle X_i\rangle_{i\in N} \in \Pi(M)} \min_{j \in N}  \valu_i(X_j).
$$

Intuitively, when allocating items to $n$ agents, 
each agent should get an allocation with a value that is $1/n$ of the total value they have for all the items.
Since the items are not divisible, this proportionality requirement may be not achievable for the agents. 
In view of this, $\MMS_i$ can be viewed as a relaxed lower bound on the value that agent $i$ hopes for 
if she has the chance to partition the items into $n$ bundles and every other agent adversarially chooses a bundle before $i$.
Next, we generalize the classical MMS notion to
the setting with asymmetric agents. 

\begin{definition}[Weighted MMS]
Given any chore allocation instance $\cI=(N,M,\bm{s},\bm{V})$, for every agent $i \in N$,
the {\em weighted maxmin share (WMMS)} value of $i$ is defined as:
\begin{align*}
\WMMS_i(\cI) = &\max_{\langle X_i\rangle_{i\in N} \in \Pi(M)} \min_{j \in N}  \valu_i(X_j) \frac{s_i}{s_j}. \label{def:2}
\end{align*}
Any partition achieves $\WMMS_i(\cI)$ is called a {\em P-$i$ partition}.
\end{definition}

When the instance $\cI$ is clear from the context, we may use $\WMMS_i$ for short.
The definition above for WMMS fairness is exactly the same as that of WMMS as formalized by \cite{FHG+17a} for the case of goods except that the entitlement $e_i$ of an agent $i$ is replaced by her share $s_i$. As mentioned in the introduction, whereas a higher entitlement for goods is desirable for an agent, a higher share for chores is undesirable for the agent.

We call an allocation WMMS if the value of the allocation to each agent $i$ is worth at least $\WMMS_i$ to her.
Similarly, an allocation is called $\alpha$-WMMS, if the total 
value of items allocated to each agent $i$ is at least ${\alpha} \WMMS_i$ for $\alpha \ge 1$. 

Note that when all shares are equal, WMMS coincides with MMS fairness so it is a proper generalization of MMS. Secondly, we spell out an insight that also provides justification for the WMMS concept that was defined by \cite{FHG+17a}.
We note that when the items are \emph{divisible}, then $\WMMS_i=s_i\valu_i(M)$.
Hence, for divisible chores, WMMS fairness also implies a natural generalization of proportionality that takes into account the shares of agents. 
We call the latter requirement as \emph{weighted proportionality}.

In the following, we define some more notation that will be used in the paper.
Given a chore allocation instance $\cI=(N,M,\bm{s},\bm{V})$, for any agent $i$ 
and any partition $X=\langle X_i\rangle_{i\in N}$, let
$\W_i^{\cI}(X)=\min_{k \in N}\frac{\valu_i(X_k)}{s_k}$.
That is, $\W_i^{\cI}(X)$ is the unfairness degree of allocation $X$ to $i$.
Let $\W_i(\cI) = \max_{\langle X_1, \ldots, X_n\rangle \in \Pi(\items)} \W_i^{\cI}(X)$.
Thus $\W_i(\cI)$ is the smallest degree of unfairness
and $\WMMS_i(\cI)=s_i\W_i(\cI)$. 
Moreover, we have the following simple properties.

\begin{lemma}\label{lem:wmms:bound}
Given any instance $\cI=(N,M,\bm{s},\bm{V})$, for any $i\in N$,
$\W_i(\cI)\leq -1$, and $\WMMS_i(\cI) \leq -s_i$.
\end{lemma}

\begin{proof}
Note that for any agent $i$ and any allocation $\langle X_i\rangle_{i\in N}$,
$\sum_{k\in N} {V_i(X_k)\over s_k}\cdot s_k = \sum_{k\in N} V_i(X_k) = -1$,
which is the weighted arithmetic mean of the terms ${V_i(X_k)\over s_k}$, with weights $s_k$ (whose sum is 1). 
As the mean equals $-1$, the smallest must be at most $-1$ and $\W_i(\cI) = \min_{k\in N} {V_i(X_k)\over s_k} \leq -1$.
\end{proof}

Next we show a simple algorithm, $\mathsf{Naive}$,
which returns an $n$-$\WMMS$ allocation.
Algorithm $\mathsf{Naive}$ produces 
an allocation that allocates all of the items to 
a single agent who has the highest share (ties are broken arbitrarily). 

\begin{lemma}
Let $\cI=(N,M,\bm{s},\bm{V})$ be any chore allocation instance
and $\langle X_i\rangle_{i\in N}$ be the output of Algorithm $\mathsf{Naive}$.
Then $V_i(X_i)\geq n\WMMS_i(\cI)$ for any $i\in N$.
\end{lemma}

\begin{proof}
Let $i^*$ be the agent who has the largest share, thus $s_{i^*}\geq \frac{1}{n}$.
It is easy to see that for any agent $i\neq i^*$, 
$V_i(X_i)=0$, which is trivially at least as large as $n\WMMS_i(\cI)$.
By Lemma \ref{lem:wmms:bound}, $\WMMS_{i^{*}}(\cI) \leq -s_{i^*} \leq -\frac{1}{n}$.
Accordingly, $V_{i^*}(M) \geq n \WMMS_{i^*}(\cI)$.
\end{proof}

We present the following example to provide additional intuition of WMMS 
and our notation. 

\paragraph{Example}
Let $\cI=(N,M,\bm{s},\bm{V})$ be a chore allocation instance,
where $N=\{1,2\}$, $M=\{1,2,3,4\}$ and the agents' shares and valuations are shown in Table \ref{example1}.
\begin{table}[h]
\begin{center}
\begin{tabular}{cc|cccc}
&&	& Chores \\
Agent & Share &1 & 2 & 3 & 4 \\
\hline
1 & $\frac{1}{4}$ & $-\frac{1}{4}$ & $-\frac{1}{4}$ & $-\frac{1}{4}$ & $-\frac{1}{4}$  \\
2 & $\frac{3}{4}$ &  $-\frac{3}{8}$ & $-\frac{3}{8}$ & $-\frac{1}{8}$ & $-\frac{1}{8}$
\end{tabular}
\end{center}
\caption{An Example of a Chore Allocation Setting.}
\label{example1}
\end{table}%


In this instance, for agent 1, allocating one of the four chores to herself and 
the remaining three chores to agent 2 is an exact weighted proportional allocation with respect to valuation $V_1$.
Then $\W_1(\cI)=-1$, $\F_1(\cI)=1$ and $\WMMS_1(\cI)=-\frac{1}{4}$.

Similarly, for agent 2, allocating chores $\{1,2\}$ to agent 2 and chores $\{3,4\}$ to agent 1 is 
an exact weighted proportional allocation with respect to valuation $V_2$.
Thus $\W_2(\cI)=-1$, $\F_2(\cI)=1$ and $\WMMS_2(\cI)=-\frac{3}{4}$.
Note that this allocation is bad for agent 1 since $V_1(\{3,4\})=-\frac{1}{2}<\WMMS_1(\cI)$.

However, one of the weighted proportional allocations to agent~1, e.g., $X_1=\{1\}$ and $X_2=\{2,3,4\}$, satisfies both of $\WMMS_{1}(\cI)$ and $\WMMS_{2}(\cI)$,
since $V_1(X_1)=-\frac{1}{4} \geq \WMMS_1(\cI)$ and $V_2(X_2)=-\frac{5}{8}\geq \WMMS_2(\cI)$.
Therefore, $\langle X_1, X_2 \rangle$ is a WMMS allocation.

\section{Optimal WMMS Fairness}

It is well known that for symmetric agents, no matter the items are goods or chores, an MMS allocation always exists for the 2-agent case. 
But for asymmetric agents, we note that an exact WMMS allocation may not exist even when there are only two agents. 
Indeed, by the following lemma, we see that the lower bound of the problem is at least $\frac{4}{3}$, 
which means that there is no allocation that can guarantee each agent's value to be greater than $\frac{4}{3}\WMMS_{i}(\cI)$ for every $i\in N$.

\begin{lemma}
\label{thm:2agent:lb}
In the chore allocation problem,
any algorithm has an approximation ratio of at least $\frac{4}{3}$ for WMMS fairness.
\end{lemma}

 \begin{proof}
 In the following we construct an instance $\cI=(N,M,\bm{s},\bm{V})$ with $N=\{1,2\}$, $M=\{1,2\}$
 and the shares and valuations are shown in Table \ref{table:lowerbound}.
 
 \begin{table}[htbp]
 \begin{center}
 \begin{tabular}{c|c|cc}
 	&&	Items \\
 \mbox{Agent} & \mbox{Share}  & 1 & 2\\
 \hline
 1 & $\frac{3}{4}$ & $-\frac{3}{4}$ & $-\frac{1}{4}$\\
 2 & $ \frac{1}{4}$ & $-\frac{1}{2}$ & $-\frac{1}{2}$
 \end{tabular}
 \end{center}
 \caption{Instance to establish the $\frac{4}{3}$ lower bound for 2 agents.}
 \label{table:lowerbound}
 \end{table}%

 We first note that for agent 1, the unique P-$1$ partition is  $X=\langle X_{1},X_{2}\rangle$ with $X_1 =\{1\}$ and $X_2 =\{2\}$,
 since $W^{\cI}_1(X)=\min\left\{\dfrac{-\frac{3}{4}}{\frac{3}{4}},\dfrac{-\frac{1}{4}}{\frac{1}{4}}\right\}=-1$,
 which is the largest among all possible allocations.
 Accordingly, $\WMMS_1(\cI)=\frac{3}{4}\times (-1)= -\frac{3}{4}$.
 For agent 2, to maximize $\W^{\cI}_{2}$,
 the only way is to set $X'=\langle X'_{1},X'_{2}\rangle$ with $X'_1=\{1,2\}$ and $X'_2=\emptyset$,
 since $\W^{\cI}_2(X')=\min\left\{\dfrac{-1}{\frac{3}{4}},\dfrac{0}{\frac{1}{4}}\right\} = -\frac{4}{3}$,
 which is the largest among all possible allocations.
 Accordingly, $\WMMS_2(\cI)=\frac{1}{4}\times (-\frac{4}{3})= -\frac{1}{3}$.

 However, $X$ is bad to agent 2, since $V_{2}(X_{2}) = -\frac{1}{2} < \WMMS_2(\cI)$ and
 $X'$ is bad to agent 1, since $V_{1}(X'_{1}) = -1 < \WMMS_1(\cI)$.
 Therefore the best tradeoff to satisfy the two agents simultaneously would be allocation $X'$,
 since $V_{2}(X_{2})= \frac{3}{2}\WMMS_2(\cI)$ and $V_{1}(X'_{1}) = \frac{4}{3} \WMMS_1(\cI)$.
 Indeed, we need to take all possible allocations into consideration, but it is easy to see that all other allocations can only be worse.

 Thus, no algorithm could provide an allocation with each agent $i$'s value being strictly larger than $\frac{4}{3}\WMMS_i(\cI)$,
 which finishes the proof of Lemma~\ref{thm:2agent:lb}.
 \end{proof}

Accordingly, it is natural to consider a relaxed version of WMMS, {\em optimal WMMS (OWMMS) fairness}, which is similar to the one  introduced in \cite{ARSW17a}.
\begin{definition}[Optimal WMMS]\label{def:opt:wmms}
Let $\cI=(N,M,\bm{s},\bm{V})$ be a chore allocation instance. 
The {\em optimal WMMS (OWMMS) ratio} $\alpha^*$ is defined as the minimal $\alpha\in[1,\infty)$ for which an $\alpha$-WMMS allocation always exists.
Let $\OWMMS_{i}(\cI)=\alpha^{*}\WMMS_{i}$ for any $i\in N$.
A partition $X=\langle X_1, \ldots, X_n\rangle$ is called an OWMMS allocation, 
if $V_i(X_i) \geq \OWMMS_{i}(\cI)$ for all $i\in N$. 
\end{definition}

It is easy to see that $\WMMS_i (\cI) \geq \OWMMS_i (\cI)$ for any instance $\cI$ and any agent $i$.
For any partition $X=\langle X_1, \ldots, X_n\rangle$, if  
$V_i(X_i) \geq c\cdot\OWMMS_i (\cI)\mbox{ for all $i\in N$},$
then $X$ is called $c$-approximation to the OWMMS allocation.

\section{Approximation Algorithms}

For the case of goods allocation, the greedy {\em round robin} algorithm considered by \cite{FHG+17a} gives the best guarantee (of $n$-approximation for goods).
Interestingly, the same algorithm was proved to provide a 2-approximation for MMS allocation of chores when agents are symmetric \cite{ARSW17a}.
However, when agents have different shares, such an algorithm can be arbitrarily poor.
We provide a bad example in the appendix, 
where we also show that some natural attempts to `fix' the bad performance of the greedy algorithm do not help.

In the following, we give our polynomial-time $(4+\epsilon)$-approximation algorithm.
That is, for any $\epsilon>0$, 
it returns an allocation $\langle X_i\rangle_{i\in N}$ such that
for any agent $i$, $V_i(X_i)\geq (4+\epsilon)\OWMMS_i$.
In order to present the main algorithm,
we first present a polynomial-time algorithm which guarantees each agent $i$'s value to be at least $2\WMMS_{i}$,
when all of the agents have an identical valuation.

\subsection{Identical Valuation}
\label{sec:4:IFV}
When all agents have an identical valuation,
we show the algorithm, $\mathsf{EgalGreedy}$ defined in Algorithm \ref{alg:WMMS:identical},
is a 2-approximation to an exact $\WMMS$ allocation.

\begin{algorithm}[htbp]
  \caption{\hspace{-3pt} $\mathsf{EgalGreedy}$ - An Algorithm for Identical Valuations}
 \label{alg:WMMS:identical}
  \begin{algorithmic}[1]
	  \footnotesize
\REQUIRE  Chore allocation instance $(N,M,\bm{s},V)$
\ENSURE Allocation $X=\langle X_1, \ldots, X_n\rangle$.
\STATE Initially, $X_{i} = \emptyset$ for all $i\in N$.
\STATE Order all chores from the lowest value 
to the highest value such that $V^{1} \leq V^{2}\leq \cdots \leq V^{m}$. \label{step:greedy:3}

\FOR{$j=1$ to $m$}
\STATE $i^*\in \argmax\limits_{i\in N} \dfrac{V(X_i \cup \{j\})}{s_i}$; \label{step:greedy:1}
\STATE $X_{i^*}=X_{i^*}\cup\{j\}$.\label{step:greedy:2}
\ENDFOR

\RETURN Allocation $X$.
\end{algorithmic}
\end{algorithm}

The next lemma relies on a connection to the parallel processors scheduling problem. 
In this problem, there is a set of jobs and a set of processors. 
Each job has to be processed exactly once on exactly one processor. 
Processors may have different speeds \cite{gonzalez1977bounds,friesen1987tighter}.
The problem specifies the time required to process a given job on a given machine. Typically, the goal of scheduling problems is to
find an assignment of the jobs such that the longest finishing time (i.e., makespan) is minimized.
A detailed survey of this line of work can be found in \cite{pinedo2016scheduling}.
We prove the following Lemma \ref{claim:approx:wmms} in the appendix.

\begin{lemma}\label{claim:approx:wmms}
For any chore allocation instance $\cI=(N,M,\bm{s},V)$, where all agents have the identical valuation $V$,
let $\langle X_i\rangle_{i\in N}$ be the allocation outputted by $\mathsf{EgalGreedy}$.
We have $V(X_i) \geq 2\WMMS_i(\cI)$ for any $i\in N$.
\end{lemma}

One may suspect that a natural generalization of $\mathsf{EgalGreedy}$ to the case that 
agents have different valuations may work well. Unfortunately, 
in the appendix, we provide an example that such an algorithm cannot have any constant approximation ratio.

\subsection{General Valuations}
Now we are ready to study the general case when agents may have different valuations.
For any chore allocation instance $\cI=(N,M,\bm{s},\bm{V})$,
let variable $\alpha$ represent the $\WMMS$ ratio,
and variable $x_{ij}\in \{0,1\}$ represent whether agent $i$ gets item $j$.
Let $\bm{x}=(x_{ij})_{i\in N, j\in M}$.
Then the problem of computing its OWMMS ratio and an OWMMS allocation can be formalized as the following integer program.

$$
\cI\cP:\begin{array}{ll}
  \min & \quad \alpha  \\
 \mbox{s.t.} & \ \left\{ \begin{array}{ll}
 \sum_{j\in M} V_{ij}x_{ij}\geq \alpha \WMMS_i(\cI), \  & \  \forall i \in N\\
 \sum_{i\in N}x_{ij} = 1, \  & \  \forall j \in M\\
 x_{ij} \in \{0,1\}, \  & \  \forall i \in N,  j \in M \\
 \alpha \geq 1.  \\
\end{array} \right.
\end{array}
$$

To solve $\cI\cP$, in what follows, we first prove a key technical lemma by using the rounding technique introduced by \cite{lenstra1990approximation},
which gives us the tool to round a fractional assignment to an integer assignment.

\begin{lemma}\label{lem:round}
Let $(N,M,\bm{s},\bm{V})$ be any chore allocation instance, $\bm{w}=(w_{1},w_{2},\cdots, w_{n})\in (\bR_-)^N$, and $\bm{t}=(t_1, t_{2},\cdots,t_{n})\in (\bR_-)^N$.
Denote by $M_i=\{j\in M | V_{ij}\geq t_i\}$ and $N_j=\{i\in N | j\in M_i\}$.
If the following linear program
$$
\cP:\begin{array}{ll}
 & \ \left\{ \begin{array}{ll}
 \sum_{j\in M_i} V_{ij}x_{ij}\geq w_i, \  & \  \forall i \in N\\
 \sum_{i\in N_j}x_{ij} = 1, \  & \  \forall j \in M\\
 x_{ij} \geq 0, \  & \  \forall i \in N,  j \in M_i \\
\end{array} \right.
\end{array}
$$
has a feasible solution, then any extreme point $\tilde{x}$ of this polytope (defining the solution space) can be rounded to
a feasible solution $\bar{x}$ of the integer program
$$
\cP':\begin{array}{ll}
 & \ \left\{ \begin{array}{ll}
 \sum_{j\in M_i} V_{ij}x_{ij}\geq w_i + t_i, \  & \  \forall i \in N\\
 \sum_{i\in N_j}x_{ij} = 1, \  & \  \forall j \in M\\
 x_{ij} \in \{0,1\}, \  & \  \forall i \in N,  j \in M_i. \\
\end{array} \right.
\end{array}
$$
\end{lemma}

 \begin{proof}
 Let $\tilde{x}$ be an extreme point of the polytope defined by $\cP$.
 Then $\tilde{x}$ contains at most $m+n$ nonzero variables due to the number of constraints in $\cP$.
 We construct a bipartite graph via $\tilde{x}$,
 $G=(N,M,E)$,
 where $N$ is the set of agents, $M$ is the set of chores and
 $E=\{(i,j) | \tilde{x}_{ij}>0, i\in N, j\in M\}$.
 Using the same argument with the proof of Theorem 1 in \cite{lenstra1990approximation},
 we know that $G$ is a pseudoforest,
 i.e., each connected component of $G$ is a tree or a tree plus one additional edge.

 Next, we round $\tilde{x}$ to $\bar{x}$.
 In any connected component $C$ of $G$,
 whenever there is a chore $j$ whose degree is 1, then it must be that for some $i$, $\tilde{x}_{ij}=1$.
 Then set $\bar{x}_{ij}=1$ and delete this chore from $C$.
 Denote by $C'$ the remained graph. Note that in $C'$, any remaining chore has a degree at least 2.
 Therefore $C'$ must contain a matching which covers all chores,
 due to the fact that $C'$ is a tree or a tree plus one additional edge.
 According to this matching, if $(i,j)$ is matched, set $\bar{x}_{ij}=1$; otherwise, set $\bar{x}_{ij}$ to be 0.
 
 Next, it suffices to verify that $\bar{x}$ is a feasible solution of $\cP'$.
 For each chore $j$, $\bar{x}$ assigns it to exactly one agent.
 Thus, $\sum_{i\in N_j}\bar{x}_{ij} = 1$ for any $j\in M$.
 For each agent $i\in N$, there is at most 1 chore $j$ such that $\tilde{x}_{ij}$
 is increased to 1.
 Since $0 \ge V_{ij}\geq t_i$,
 $$\sum_{j\in M_i} V_{ij}\bar{x}_{ij} \geq \sum_{j\in M_i} V_{ij}\tilde{x}_{ij} + t_i \geq w_i + t_i,$$
 which completes the proof of Lemma \ref{lem:round}.
 \end{proof}

Note that, solving the optimal $\alpha$ for integer program $\cI\cP$
is equivalent to finding the minimum value of $\alpha$ such that $\cI\cP$ has a feasible integer solution $\bm{x}$.
However, $\cP$ is not the relaxation of $\cI\cP$ 
since in $\cP$, there is not a variable $x_{ij}$ for which $V_{ij}<t_{i}$. 
Equivalently, we can add these variables to $\cP$ and set them to zero. 
In the following, we discuss the relationship between the solutions of $\cI\cP$, $\cP$ and $\cP'$.

Recall that $\alpha^{*}$ is the OWMMS ratio, which is also the optimal value of $\cI\cP$.
Let $$c^{*}=\min\{c\in \bR^{+}_0 | \cP \mbox{ has a feasible solution with }$$
$$t_{i}=w_{i}=c\WMMS_{i}(\cI) \mbox{ for all } i\in N\}.$$
Note that $c^*$ always exists as $c=n$ is always feasible by Algorithm $\mathsf{Naive}$.
Moreover, although $\cP$ is not the relaxation of $\cI\cP$, $c^{*}$ is still a lower bound of $\alpha^{*}$.

\begin{lemma}\label{lem:lbAlpha}
$\alpha^{*} \geq c^{*}$.
\end{lemma}

Lemma \ref{lem:lbAlpha} shows that to approximate $\alpha^{*}$, 
it suffices to find a feasible solution of $\cI\cP$ whose value is a good approximation to $c^{*}$. 
Next, we show that a feasible solution of $\cP'$ is naturally a feasible solution of $\cI\cP$.

\begin{lemma}
\label{lem:R'toIR}
Let $c\in \bR^{+}_0$.
If $\bm{x}$ is a feasible solution of $\cP'$ with $t_{i}=w_{i}=c\WMMS_{i}(\cI)$ for all $i\in N$,
then $(\bm{x}, 2c)$ is a feasible solution of $\cI\cP$.
\end{lemma}

Both Lemmas \ref{lem:lbAlpha} and \ref{lem:R'toIR} are proved in the appendix.
Before we show our main algorithm, let us discuss the following intuitive procedure.
First, compute $c^{*}$ and its corresponding fractional allocation $\tilde{\bm{x}}$.
Then use Lemma \ref{lem:round} to round $\tilde{\bm{x}}$ to an integer solution $\bar{\bm{x}}$.
By Lemma \ref{lem:R'toIR}, $\bar{\bm{x}}$ is also a feasible solution of $\cI\cP$.
Let $X=\langle X_i\rangle_{i\in N}$ be the final allocation, where $X_{i}=\{j\in M | \bar{x}_{ij}=1\}$ for every $i\in N$.
Thus, 
$$V_{i}(X_{i}) \geq 2c^{*}\WMMS_{i}(\cI)\geq 2\alpha^{*}\WMMS_{i}(\cI),$$
where the first inequality is by Lemma \ref{lem:round} and the second inequality is by Lemma \ref{lem:lbAlpha}.
That is, $X$ is a 2-approximation to the optimal $\WMMS$ allocation.

However, there are two computational issues with respect to the procedure above:
(1)~The computation of $\WMMS_i(\cI)$ may need exponential time\footnote{
The computation of $\WMMS_{i}$ is NP-hard, even when $n = 2$ and $s_{1} = s_{2} = \frac{1}{2}$, 
via a reduction from the Integer Partition Problem.
};
(2)~Even if we know all the $\WMMS_i(\cI)$'s, there is a problem of computing $c^{*}$. 

To resolve (1), we use Algorithm $\mathsf{EgalGreedy}$ 
to compute an approximate value $\WMMS'_{i}$ for each $\WMMS_{i}(\cI)$,
where $\WMMS'_{i} \geq 2\WMMS_{i}(\cI)$.
Then we replace all $\WMMS_{i}(\cI)$ by $\WMMS'_{i}$ in above procedure.   

To resolve (2), we use binary search to find a near optimal value of $c^{*}$.
Initially, we first use Algorithm $\mathsf{Naive}$ to get an upper bound $n$ of $\alpha$ and 1 is a trivial lower bound. 
Let $\delta>0$ be the desired precision.
Denote by $u$ and $l$ the current upper and lower bounds, respectively. 
Set $c=\frac{u+l}{2}$ and $w_i=t_i=c\WMMS'_i$, and 
check if $\cP$ has a feasible solution.
If $\cP$ has a feasible solution, reset $u$ to be $\frac{u+l}{2}$;
Otherwise reset $l=\frac{u+l}{2}$.
Repeat this process until $u-l\leq \delta$.

We formally describe the algorithm described above as Algorithm~\ref{alg:linearProgramming}, denoted by $\mathsf{LinPro}$.

\begin{algorithm}[htbp]
  \caption{\hspace{-3pt} $\mathsf{LinPro}$ - An Algorithm for General Valuations}
 \label{alg:linearProgramming}
  \begin{algorithmic}[1]
	  	  \footnotesize
\REQUIRE  Chore allocation instance $\cI=(N,M,\bm{s},\bm{V})$ and $\epsilon>0$.
\ENSURE Allocation $X=\langle X_1, \ldots, X_n\rangle$
\STATE Initially, $X_{i} = \emptyset$ for all $i\in N$.
\STATE For each $i\in N$, run Algorithm $\mathsf{EgalGreedy}$ on instance $\cI$ and obtain allocation $X^i=\langle X^{i}_1, X^{i}_2, \ldots, X^{i}_n\rangle$. 
 \STATE Set $\WMMS'_i = V_i(X^i_i)$ for all $i\in N$. 
\STATE Let $u=n$ and $l=1$. \% $n$ is the upper bound by $\mathsf{Naive}$.
 \WHILE{$u-l>\frac{\epsilon}{4}$}
   \STATE Set $c=\frac{u+l}{2}$.
   \STATE Check if $\cP$ has a feasible solution by setting $w_i=t_i=c\WMMS'_i$ for all $i\in N$. 
   \IF{$\cP$ has a feasible solution}
   \STATE Reset $u=c$.
   \ELSE
   \STATE Reset $l=c$.
   \ENDIF
 \ENDWHILE
\STATE Set $w_i=t_i=u\WMMS'_i$ for all $i\in N$ and compute an extreme point $\tilde{\bm{x}}$ of $\cP$. \label{step:L:endofbinary}
\STATE Use Lemma \ref{lem:round} to round $\tilde{\bm{x}}$ to $\bar{\bm{x}}$. 
\STATE Set $X_i=\{j\in M | \bar{x}_{ij} = 1\}$ for all $i\in N$.
\RETURN Allocation $X$.
\end{algorithmic}
\end{algorithm}

\begin{theorem}
Given any chore allocation instance $\cI=(N,M,\bm{s},\bm{V})$ with  $\alpha^*$ being its OWMMS ratio.
For any $\epsilon>0$, Algorithm $\mathsf{LinPro}$ runs in polynomial time (for any number of agents) 
and returns an allocation $\langle X_i\rangle_{i\in N}$ such that
for any agent $i$, $V_i(X_i)\geq (4+\epsilon)\OWMMS_i(\cI)$.
\end{theorem}

\begin{proof}
By Lemma \ref{lem:R'toIR}, $X$ is a feasible solution of $\cI\cP$.
At Step \ref{step:L:endofbinary}, as $l\leq c^{*}\leq u$ and $u-l\leq \frac{\epsilon}{4}$, we have $u\leq c^*+\frac{\epsilon}{4}$.
Thus, 
\begin{align*}
V_i(X_i)  \geq &  2u\WMMS'_i \geq 2(c^{*}+\frac{\epsilon}{4})\WMMS'_i  \\
\geq& 4(c^{*}+\frac{\epsilon}{4})\WMMS_i(\cI) \geq  4(\alpha^{*}+\frac{\epsilon}{4})\WMMS_i(\cI) \\
\geq& (4+\epsilon)\OWMMS_i(\cI),
\end{align*}
where the first inequality is by Lemma \ref{lem:round},
the second inequality is by Lemma \ref{claim:approx:wmms},
and the last inequality is by Lemma \ref{lem:lbAlpha}.

As $\mathsf{LinPro}$ requires us to run $\mathsf{EgalGreedy}$ and solve at most $O(\log (\frac{n}{\epsilon}))$  numbers of (polynomial-sized) linear program, 
$\mathsf{LinPro}$ runs in polynomial time. 
\end{proof}

Note that the role of $\mathsf{EgalGreedy}$ in $\mathsf{LinPro}$ can be replaced by other (polynomial-time) approximation algorithms (such the PTAS in \cite{Hochbaum1988a})
and the approximation ratio (Lemma \ref{claim:approx:wmms}) is improved accordingly. 

\section{Restricted Cases}

In this section, we consider two important restricted cases: (1) two agents and 
(2) all agents have binary valuations (in which case agents have value 0 or -1 for each item).

\subsection{WMMS for Two Agents}
\label{sec:2agents}
Given any instance  $\cI=(N,M,\bm{s},\bm{V})$ with $N=\{1,2\}$,
we prove that it is always possible to guarantee each agent $i$'s value to be at least $\frac{3}{2}\WMMS_{i}(\cI)$.
Thus, by Lemma \ref{thm:2agent:lb}, the OWMMS ratio $\alpha^{*}$ for the 2-agent case is within $[\frac{4}{3}, \frac{3}{2}]$.

Divide-and-choose algorithms are widely studied in the literature, especially for the case of two agents.
Roughly speaking, the algorithm starts by letting one of the agents divide the whole items (either goods or chores) into two bundles,
and the other agent chooses one from the two bundles.
Such an algorithm gives an exact MMS allocation for symmetric agents and
a 2-WMMS allocation for asymmetric agents (the agent with smaller entitlement divides and the other agent chooses) when the items are goods.
However, it is not hard to see that generic divide-and-choose algorithms could be arbitrarily bad when the items are chores.

In the following, we show that with some modification, a divide-and-choose style algorithm, $\mathsf{DivCho}$ (defined in Algorithm \ref{alg:WMMS:2agents}), 
works well and guarantees each agent $i$'s value to be at least $\frac{3}{2}\WMMS_{i}$.
Without loss of generality, assume $s_1\leq s_2$.

\begin{algorithm}[htbp]
  \caption{\hspace{-3pt} $\mathsf{DivCho}$ -An Algorithm for the 2-Agent Case}
 \label{alg:WMMS:2agents}
  \begin{algorithmic}[1]
	  	  \footnotesize
\REQUIRE  Chore allocation instance $\cI=(N,M,\bm{s}, \bm{V})$ with $N=\{1,2\}$.
\ENSURE Allocation $X=\langle X_1, X_2\rangle$.
\STATE Initially, set $X_{i}=\emptyset$ for both $i\in N$.
\STATE If $s_1\leq \frac{1}{3}$ and $s_2\geq \frac{2}{3}$, set $X_1=\emptyset$ and $X_2=M$. Go to Step \ref{step:dc:output}.
\STATE Let agent 2 partition $M$ into $A_1$ and $A_2$ according to a P-2 partition with respect to $\WMMS_2(\cI)$.
\STATE Let agent 1 select his favorite bundle from $A_1$ and $A_2$.
Denote by $X_1$ the one chosen by agent 1 and by $X_2$ the one left for Agent~2.
\STATE $X=\langle X_1, X_2\rangle$.\label{step:dc:output}
\RETURN Allocation $X$.
\end{algorithmic}
\end{algorithm}

\vspace{-5mm}

\begin{theorem}
\label{thm:divcho}
Let $\cI=(N,M,\bm{s}, \bm{V})$ with $N=\{1,2\}$,
and $X=\langle X_1, X_2\rangle$ be the output of Algorithm $\mathsf{DivCho}$ on $\cI$.
Then, for any agent $i\in N$, $V_i(X_i)\geq \frac{3}{2}\WMMS_i(\cI)\geq  \frac{3}{2}\OWMMS_i (\cI)$.
\end{theorem}

 \begin{proof}
 If $s_1\leq \frac{1}{3}$ and $s_2\geq \frac{2}{3}$,
 Algorithm $\mathsf{DivCho}$ allocates all chores to agent 2.
 Thus $V_{1}(X_{1})=0$ and $V_{2}(X_{2})=-1$, where agent 1 is trivially satisfied as $V_1(X_1)\geq \frac{3}{2}\WMMS_1(\cI)$.
 By Lemma \ref{lem:wmms:bound}, we have $\WMMS_2(\cI) \leq -s_2 \leq -\frac{2}{3}$.
 Accordingly, $V_2(X_2)\geq \frac{3}{2}\WMMS_2(\cI)$.

 As we assume that $s_2 \ge s_1$,
 our last case is to consider $\frac{1}{2} \leq s_{2} \leq \frac{2}{3}$. 
 By the definition of  $\WMMS_2(\cI) = s_{2}\cdot\min\{\frac{V_{2}(A_{1})}{s_{1}},\frac{V_{2}(A_{2})}{s_{2}}\}$,
 $V_2(A_1)\geq V_2(A_2) \geq \WMMS_2(\cI)$.
 As a result, no matter which allocation agent 2
 eventually receives after the divide-and-choose procedure,
 the value of the allocation will always be at least as much as $\WMMS_2$.
 For agent 1, since $V(X_1) + V(X_2) = -1$ by assumption and
 he selects his favorite allocation $X_1$,
 $V_1(X_1)\geq -\frac{1}{2}$.
 By Lemma \ref{lem:wmms:bound} and the fact that $s_1> \frac{1}{3}$, $\WMMS_1(\cI) < -\frac{1}{3}$.
 Therefore, $V_1(X_1) > \frac{3}{2}\WMMS_1(\cI)$.
 \end{proof}




\subsection{Binary Valuations}

\label{sec:binary}

In this section, we study the case with any number of agents, but every agent's valuation is binary: $V_{ij}\in \{0,-1\}$ for all $i\in N$ and $j\in M$. 
Note that, throughout this section, we do not impose normalization for ease of exposition. 
As will be clear later, for this case, we show that it is always possible to guarantee each agent $i$'s
value to be at least $\WMMS_{i}$, i.e., the optimal $\WMMS$ ratio for binary valuation case is exactly 1.

We first prove the following lemma. 

\begin{lemma}\label{claim:same valuation}
Let $\cI=(N,M,\bm{s},V)$ be a chore allocation instance where all agents have an identical valuation $V$.
If $V$ is uniform, (i.e. $V(S)=-|S|$ for any $S\subseteq M$), an exact $\WMMS$ allocation can be computed in polynomial time.
\end{lemma}


 \begin{proof}

 It suffices to show if $V$ is uniform,  Algorithm $\mathsf{EgalGreedy}$ 
 returns an exact $\WMMS$ allocation.  
 Suppose $X=\langle X_i\rangle_{i\in N}$ is the output of $\mathsf{EgalGreedy}$ with respect to $V$.
 Recall $W(X) = \min_{k\in N}\frac{V(X_k)}{s_k}= \min_{k\in N}\frac{-|X_k|}{s_k}$. 
 Let $S =\{k\in N | \frac{-|X_k|}{s_k} = W(X) \}$ be the set of indices where the minimum is obtained.
 In the following we show $W=W(X)$.
 Note that $S\neq \emptyset$.
 If $X$ is not an optimal partition, then there is a partition $X^{*}$ such that $W(X^{*})>W(X)$.

 Thus, every $k\in S$, $X_{k}$ has to contain a smaller number of chores compared with $X^{*}_{k}$.
 Accordingly, for some $t\in N\backslash S$, $X_{t}$ has to contain more chores than $X^{*}_{t}$, i.e., $V(X_{t}^{*}) \leq -|X_t|-1$.
 If $\frac{-|X_t|-1}{s_t} \leq \W(X)$, $\W(X^{*})$ cannot be larger than $W(X)$.
 Thus, $\frac{-|X_t|-1}{s_t} > \W(X)$.
 But this is a contradiction with the fact that $\mathsf{EgalGreedy}$ always
 allocates greedily, (i.e., Step \ref{step:greedy:1} of $\mathsf{EgalGreedy}$),
 since the last chore cannot be allocated to $X_{k}$ for $k\in S$ instead of $X_{t}$.
 That is $\W=\W(X)$.

 Therefore, $\frac{V(X_{i})}{s_{i}}\geq \W(X)=\W$ and $V(X_{i})\geq s_{i}\W=\WMMS_{i}$ for any $i\in N$, which competes the proof.
 \end{proof}

Thus, by allocating all chores for which some agent has zero value to one such agent, 
we are left with only the chores for which all agents have value -1.
As the modified instance if uniform, by Lemma~\ref{claim:same valuation}, we have the following theorem.  
\begin{theorem}
For any binary valuation case, a $\WMMS$ allocation exists and can be found efficiently.
\end{theorem}


%
%


\section{Conclusions}
We initiated the study on chore allocation with asymmetric agents.
We show that many widely studied greedy algorithms in the literature performs badly and 
even for the 2-agent case an exact WMMS allocation may not exist.
We then presented a constant approximation polynomial time algorithm for OWMMS allocations,
and several algorithmic results for the case of identical utilities, binary utilities, and for 2 agents. 
Finding a  stronger lower bound for WMMS allocations for any number of agents remains an open problem.

\section*{Acknowledgements}
This work is partially supported by NSF CAREER Award No. 1553385.
Haris Aziz is supported by a Scientia Fellowship.




\newpage
\appendix

\section{Some Commonly Used Greedy Algorithms}

In this section, we first show that the greedy round robin algorithm considered by \cite{FHG+17a} which gives the best guarantee (of $n$-approximation for goods) can be arbitrarily poor for the case of chores. This is surprising because the algorithm only uses ordinal preferences and higher entitlements and higher shares correlate with more items in the goods and chores setting respectively. We also show that natural attempts to `fix' the bad performance of the greedy sequential algorithm does not help. Interestingly, the same round-robin greedy algorithm was proved to provide a 2-approximation for MMS allocation of chores when agents have the same shares~\cite{ARSW17a}.

\subsection{Round Robin}
We define the greedy algorithm $\mathsf{Round\text{-}Robin}$ as follows. 
The algorithm is oblivious to the shares of the agents. 
It is based on sequentially allocating items in a round robin manner. 
Each agent gets turns in a round robin manner to select one of her most preferred chores from all unselected chores. 

\begin{quote}
	$\mathsf{Round\text{-}Robin}$: 
	Specify an ordering of agents and let agents come in a round robin manner in the specified order and 
	pick an item that is most preferred from the unallocated items. 
	Stop when all the items have been allocated.
\end{quote}

Now we construct a bad instance.
Let $\cI=(N,M,\bm{s},V)$ be a chore allocation instance with $n$ agents and $n^2$ items,
and $s_{i}=\frac{n}{(n+1)^{n-i+1}}$ for $i\in N$.
All agents have identical valuation $V$, defined as follows.
Denote $j\in M$ by $j=kn+b$, where $0\leq k \leq n$ and $0 \leq b<n-1$,
and $V_{j} = -\frac{1}{(n+1)^{n-k+1}}$.
That is every agent has value
$$ -\frac{1}{(n+1)^{n}} \mbox{ for any item in } \{1,\cdots, n\};$$
$$  -\frac{1}{(n+1)^{n-1}} \mbox{ for any item in } \{n+1,\cdots, 2n\};$$
$$\vdots$$
$$  -\frac{1}{n+1} \mbox{ for any item in } \{n^{2}-n+1,\cdots, n^{2}\}.$$

Since $\sum_{i\in N} s_{i} \to 1$ and $V(M) = \sum_{j \in M} V_{j}\to 1$ as $n\to\infty$, 
the instance is well-defined.
Note that it is easy to see the weighted proportional allocation with respect to $V$ is to allocate all chores in $\{(i-1)n +1, \cdots, in\}$ to agent $i\in N$,
where every agent's absolute value equals to her share, thus $\WMMS_{i}(\cI) = s_{i} =-\frac{n}{(n+1)^{n-i+1}}$.

However, following the $\mathsf{Round\text{-}Robin}$ protocol,
the items selected by the agents is as follows:
for any $1\leq k \leq n$, during the $k$th round, each agent $i$ will select one item from $\{(k-1)n +1, \cdots, kn\}$.
That is all items are `uniformly' distributed among all agents such that each agent has value $-\frac{1}{n}$ for his own bundle.
Let us consider agent 1 who has the smallest share with $\WMMS_{1}(\cI)=-\frac{n}{(n+1)^{n}}$.
Since 
$$\dfrac{\frac{1}{n}}{\frac{n}{(n+1)^{n}}} \to \infty,$$
the $\mathsf{Round\text{-}Robin}$ allocation to agent 1 is arbitrarily bad.
Thus, $\mathsf{Round\text{-}Robin}$ does not have any bounded approximation guarantee for the optimal $\WMMS$ allocation.

\subsection{Multiplicative-Greedy Algorithm}
Next we consider the following greedy algorithm to modify the Round Robin algorithm. 
In contrast to $\mathsf{Round\text{-}Robin}$, the picking order of the agents changes dynamically.

 \begin{quote}
 $\mathsf{Multiplicative\text{-}Greedy}$: Let each agent's proportionality value be $V_i(M)s_i$,
 which is exactly $s_i$ if we normalize all valuations.
 Initialize allocation $X_i$ to be empty for each $i \in N$.
 Consider the agent $i$ for whom $V_i(X_i)/s_i$ is the minimum.
 In case of ties, choose the agent with the largest $s_i$. If there is still a tie, break tie lexicographically.
 Let the agent $i$ select her most preferred untaken item.
 Repeat until all items are allocated.
 \end{quote}

 Net we present a bad example to show that $\mathsf{Multiplicative\text{-}Greedy}$ cannot provide a good guarantee as well.
 We consider the following example shown in Table~\ref{table:multiplicatedgreedy:1}.
 Thus agent 2 selects item 2 and then agent 1 selects item~1.
 This allocation is arbitrarily bad to agent 1 if $0<\epsilon<1$ is sufficiently small.

 \begin{table}[htbp]
 \begin{center}
 \begin{tabular}{cc|cc}
 	&&Chores\\
 \mbox{Agent} & \mbox{Share}  & \mbox{1} & \mbox{2} \\
 \hline
 1 & $\epsilon$ & $-1+\epsilon$ & $-\epsilon$  \\
 2 & $1-\epsilon$ & $-1+\epsilon$ & $-\epsilon$
 \end{tabular}
 \end{center}
 \caption{Instance 1 on which $\mathsf{Multiplicative\text{-}Greedy}$  performs badly. }
 \label{table:multiplicatedgreedy:1}
 \end{table}%

 On the other hand, if we modify $\mathsf{Multiplicative\text{-}Greedy}$ by using the smallest share to break ties, the algorithm performs poorly on following example in Table~\ref{table:multiplicatedgreedy:2}.

 \begin{table}[htbp]
 \begin{center}
 \begin{tabular}{cc|cccc}
 	&&Chores\\
 \mbox{Agent} & \mbox{Share}  & \mbox{1} & \mbox{2} & 3 & 4 \\
 \hline
 1 & $\epsilon$ & $-\epsilon+\epsilon^2$ & $-\epsilon^2$ & $-\epsilon$ & $-1+2\epsilon$  \\
 2 & $\epsilon$ &  $-\epsilon+\epsilon^2$ & $-\epsilon^2$ & $-\epsilon$ & $-1+2\epsilon$ \\
 3 & $1-2\epsilon$ &  $-\epsilon+\epsilon^2$ & $-\epsilon^2$ & $-\epsilon$ & $-1+2\epsilon$
 \end{tabular}
 \end{center}
 \caption{Instance 2 on which $\mathsf{Multiplicative\text{-}Greedy}$  performs badly. }
 \label{table:multiplicatedgreedy:2}
 \end{table}%

 $\mathsf{Multiplicative\text{-}Greedy}$ will run as follows:
 Agent 1 selects item 2;
 Agent 2 selects item 1;
 Agent 3 selects item 3;
 At this time, $\frac{V_1(X_1)}{s_1} =\epsilon$, $\frac{V_2(X_2)}{s_2} = 1-\epsilon$ and $\frac{V_3(X_3)}{s_3} = \frac{\epsilon}{1-2\epsilon}$.
 Since $\frac{V_1(X_1)}{s_1}< \frac{V_3(X_3)}{s_3} < \frac{V_2(X_2)}{s_2}$, agent 1 need to select item 4.
 However, it is easy to see that $\WMMS_1=\WMMS_2=-\epsilon$, $\WMMS_3=-1+2\epsilon$,
 and there exists an WMMS allocation: $X'_1=\{1,2\}$, $X'_2=\{3\}$ and $X'_3=\{4\}$.
 Since for agent 1, $\frac{V_1(X_1)}{\WMMS_1} =\frac{1-2\epsilon+\epsilon^{2}}{\epsilon} \to \infty$,
 the returned allocation $X$ is arbitrarily bad to agent 1.

 \subsection{Additive-Greedy Algorithm}


 We consider another sequential allocation greedy algorithm that uses an additive criterion to decide which agent gets the turn to pick an item.
 Just like the $\mathsf{Multiplicative\text{-}Greedy}$, the picking sequence of the agents is not pre-defined and it changes according to the items that have been allocated.

 \begin{quote}
 $\mathsf{Additive\text{-}Greedy}$:
 Initialize allocation $X_i$ to be empty for each $i \in N$.
 Consider the agent $i$ for whom $s_i + V_i(X_i)$ is the maximum.
 In case of ties, choose the agent with the largest $s_i$.
 If there is still a tie, break tie lexicographically.
 Let the agent select the most preferred untaken item. Repeat until all items are selected.
 \end{quote}

 \begin{table}[htbp]
 \begin{center}
 \begin{tabular}{cc|cccc}
 &&	Chores \\
 \mbox{Agent} & \mbox{Share}  & 1 & 2 & 3 & $4,\cdots,m$ \\
 \hline
 1 & $\epsilon$ & $-1+\epsilon$ & 0 & $-\epsilon$ & 0 \\
 2 & $1-\epsilon$ & $-\epsilon+\epsilon^2$ & $-\epsilon$ & $-\epsilon^2$ & $-\epsilon^2$
 \end{tabular}
 \end{center}
 \caption{Instance on which $\mathsf{Additive\text{-}Greedy}$ performs badly. }
 \label{table:additivegreedy:1}
 \end{table}%

 Next we show this algorithm also has a bad performance. Consider the example shown in Table \ref{table:additivegreedy:1}.
 By setting $m$ and $\epsilon$ such that 
 $$(-\epsilon+\epsilon^2)+ (-\epsilon) + (m-2)(-\epsilon^2) = -1.$$
 In the beginning, agent 2 has a larger share and she selects her most preferred item,
 i.e. one from $\{3,\cdots,m\}$ since all of them are her favorite items.
 Note that agent 2 still has a larger value with respect to criterion $s_i + V_i(X_i)$, until all $\{3,\cdots,m\}$ have been selected by agent 2.
 At this time, $s_2 + V_2(X_2) = (1-\epsilon) -(m-2)\epsilon^2 = \epsilon-\epsilon^2$.
 Since $s_1 - V_1(X_1) =\epsilon > s_2 + V_2(X_2)$,
 agent 1 will be the next to select her most preferred item from $\{1,2\}$. Agent 1 will select item 2 since its value is  0.
 But this does not affect the value of $s_i - V_i(X_i),i=1,2$, and
 then agent 1 has to continue to select an item, and only item 1 remains unselected.
 Thus agent 1 will eventually get item 1.
 However, it is easy to observe that $X'_1=\{2\}$ and $X'_2=\{1,3,4\}$ is an WMMS allocation as
 $\WMMS_1=-\epsilon$, $\WMMS_2=-1+\epsilon$.
 Then the algorithm is arbitrarily bad to agent 1.

\section{$\mathsf{EgalGreedy}$ Does not Work for General Case}

%
%

Readers may suspect that algorithm $\mathsf{EgalGreedy}$ has a good performance in the general setting,
if Steps \ref{step:greedy:1} and \ref{step:greedy:2} are replaced by finding
$$i^*= \argmax\limits_{i\in N} \dfrac{V_i(X_{i} \cup \{j\})}{s_i}$$ and
$X_{i^*}=X_{i^*}\cup\{j\}$.
Unfortunately, in the following example, we will see this is not true.
Note that Step \ref{step:greedy:3} is invalid since the agents may not have same order of values, thus this step has to be skipped.
In the following we consider the case when valuations are all normalized to $-1$.
(Indeed, if the agents' valuations are not normalized to $-1$,
it is easy to see that $\mathsf{EgalGreedy}$ is arbitrarily bad.)

Let $T>c>1$ be any two constant numbers and $n$ be a sufficiently large number such that $\frac{1}{c}+\frac{n-1}{T}=1$.
Let $k=T\sqrt{\frac{2}{c}}$.
Consider the valuation functions shown in Table~\ref{table:GisBad}.

\begin{table*}[h!]
	\begin{center}
		\begin{tabular}{cc|cccccccc}
			&&	Chores \\
\mbox{Agent} & \mbox{Share}  & 1 & 2 & $\cdots$ & $k$ & $k+1$ & $\cdots$ & $n-1$ & $n$ \\
\hline
1 & $\frac{1}{c}$ & $-\frac{1}{T}$ & $-\frac{1}{T}$ & $\cdots$  & $-\frac{1}{T}$& $-\frac{1}{T}$& $\cdots$ & $-\frac{1}{T}$ & $-\frac{1}{c}$ \\
2 & $\frac{1}{T}$ & $-\frac{c}{T^{2}}$ & $-\frac{2c}{T^{2}}$ & $\cdots$ & $-\frac{kc}{T^{2}}$ & 0 & $\cdots$ & 0 & 0\\
$\vdots$ & $\vdots$ & $\vdots$ & $\vdots$\\
$n$ & $\frac{1}{T}$ & $-\frac{c}{T^{2}}$ & $-\frac{2c}{T^{2}}$ & $\cdots$ & $-\frac{kc}{T^{2}}$ & 0 & $\cdots$ & 0 & 0\\
		\end{tabular}
	\end{center}
	\caption{Algorithm $\mathsf{EgalGreedy}$ performs badly when each agent has different valuations.}
\label{table:GisBad}
\end{table*}

By the selection of $k$, each agent's value for the grand bundle goes to $-1$,
thus the example is well-defined.
Note that for any $j\leq k$, 
$$
V_{1}([1:j])=\frac{j}{T}, \mbox{ and }\dfrac{V_1([1:j])}{s_1}=-\frac{jc}{T},
$$
$$
V_{i}(\{j\})=\frac{jc}{T^{2}}, \mbox{ and }\dfrac{V_i(\{j\})}{s_i}=-\frac{jc}{T}.
$$
Thus in the first $k$ rounds of the algorithm, all the chores are allocated to agent 1.
Note that we can add some $\epsilon$ to break tie by allocating the chore to the agent with larger share.
Eventually, $[1:k]\subseteq X_{1}$ and $V_{1}(X_{1})\leq -\frac{k}{T}\approx -\sqrt{\frac{2}{c}}$.
However, it is not hard to see that $\WMMS_{1}=-\frac{1}{c}$.
Thus Algorithm $\mathsf{EgalGreedy}$ is arbitrarily bad for this example.

\section{Omitted Proofs}

\subsection{Proof of Lemma \ref{claim:approx:wmms}}


\begin{proof}
Indeed, when all agents have identical valuation,
the chore allocation problem has a similar setting with {\em scheduling on uniform processors},
where each processor may have a different speed~\cite{gonzalez1977bounds,friesen1987tighter}.
Recall that $\F_i(\cI)$ is defined as
$$\F_{i}(\cI) = \min_{\langle X_1, X_2, \ldots, X_n\rangle \in \Pi(M)} \max_{j \in N }\frac{D(X_j)}{s_j},$$
where $D=-V$. As all agents have the same value of $\F_{i}(\cI)$, we denote this value to be $\F(\cI)$.

To reduce our problem to scheduling on uniform processors, 
let every agent $i \in N$ be a processor with a speed of $s_i$.
Let every chore $j \in M$ be a job, and $D_{j}$ be the size of job $j$.
Thus, $\frac{D_{j}}{s_i}$ is $j$'s processing time if it is processed on $i$.
Then $\F(\cI)$ can be described as finding a schedule to assign every job to a processor
such that the makespan of all processors is minimized.
Therefore, the computation of $\F(\cI)$ becomes the computation of the minimum makespan of
the corresponding scheduling problem.

It is shown in \cite{gonzalez1977bounds} that Algorithm $\mathsf{EgalGreedy}$
returns a schedule whose makespan $\F'$ is at most twice of the optimal schedule's makespan $\F(\cI)$, i.e., $\F'\leq 2\F(\cI)$.
Let $X = \langle X_1, X_2, \ldots, X_n\rangle$ be the
allocation from $\mathsf{EgalGreedy}$. Thus for any agent $i\in N$,
$$\frac{D(X_i)}{s_i} \leq \F'\leq 2\F(\cI) \Rightarrow D(X_i) \leq 2s_i\F(\cI) = -2\WMMS_i(\cI).$$
That is $V(X_i) \geq 2\WMMS_i(\cI)$, which completes the proof.
\end{proof}

\subsection{Proof of Lemma \ref{lem:lbAlpha}}
%

\begin{proof}
Let $(\alpha^{*}, \bm{x})$ be an optimal solution of $\cI\cP$.
Since $\bm{x}$ is an integer solution, for any $x_{ij}=1$, it must be that $V_{ij} \geq \alpha^{*}\WMMS_{i}(\cI)$ 
as $V_{ij} \le 0$ for all $i \in N$ and $j \in M$. 
Otherwise, $V_{ij} < \alpha^{*}\WMMS_{i}(\cI)$ implies $\sum_{j\in M} V_{ij}x_{ij} < \alpha^{*} \WMMS_i(\cI)$,
which is a contradiction to that fact that $(\alpha^{*}, \bm{x})$ is a feasible solution of $\cI\cP$.
Thus, $\bm{x}$ must also be a feasible solution of $\cP$ when $t_{i}=w_{i}=\alpha^{*}\WMMS_{i}(\cI)$ for all $i\in N$,
which means $\alpha^{*} \geq c^{*}$.
\end{proof}

\subsection{Proof of Lemma \ref{lem:R'toIR}}
\begin{proof}
This is because any $x_{ij}$ with  $V_{ij}<t_{i}$ is set to be~0.
\end{proof}

\end{document}